\documentclass[10pt,conference]{IEEEtran}
\usepackage[flushleft]{threeparttable}
\usepackage{url}
\usepackage{array}
\usepackage{multirow}
\usepackage{graphicx}
\usepackage{comment}
\usepackage[dvipsnames]{xcolor}
\usepackage{xspace}
\usepackage{booktabs}
\usepackage[most]{tcolorbox}
\usepackage{balance}

\newcommand{\ie}{\textit{i.e.,}}

\definecolor{amber}{rgb}{1.0, 0.75, 0.0}

\definecolor{main}{HTML}{757677}    
\definecolor{sub}{HTML}{DEE0E3}     
\newtcolorbox{boxH}{
	colback = sub, 
	colframe = main, 
	boxrule = 0pt, 
	leftrule = 6pt 
}

\begin{document}
\def\isdoubleblindsub{1}

\newcommand{\cmcalc}{{\em qmcalc}}

\title{Broken Windows: Exploring the Applicability of a Controversial Theory on
	Code Quality}


\author{\IEEEauthorblockN{Diomidis~Spinellis\IEEEauthorrefmark{1},
		Panos~Louridas\IEEEauthorrefmark{2}, 
		Maria~Kechagia\IEEEauthorrefmark{3}, 
		Tushar~Sharma\IEEEauthorrefmark{4}}
	\IEEEauthorblockA{
		\textit{Athens University of Economics and Business}\IEEEauthorrefmark{1}\IEEEauthorrefmark{2},
		\textit{University College London}\IEEEauthorrefmark{3},
		\textit{Dalhousie University}\IEEEauthorrefmark{4}\\
		Athens, Greece\IEEEauthorrefmark{1}\IEEEauthorrefmark{2},
		London, UK\IEEEauthorrefmark{3},
		Halifax, Canada\IEEEauthorrefmark{4} \\
		dds@aueb.gr\IEEEauthorrefmark{1}, louridas@aueb.gr\IEEEauthorrefmark{2},
		m.kechagia@ucl.ac.uk\IEEEauthorrefmark{3},
		tushar@dal.ca\IEEEauthorrefmark{4}}
	}

\maketitle

\begin{abstract}
  Is the quality of existing code correlated with the quality of subsequent
  changes? According to the (controversial) broken windows theory,
  which inspired this study, disorder sets descriptive norms
  and signals behavior that further increases it. From a large code
  corpus, we examine whether code history does indeed affect the
  evolution of code quality. We examine C code quality
  metrics and Java code smells in specific files,
  and see whether subsequent commits by developers continue on that path.
  We check whether developers tailor the quality of their commits
  based on the quality of the file they commit to.
  Our results show that history matters, that developers behave
  differently depending on some aspects of the code quality they encounter, and
  that programming style inconsistency is not necessarily related to
  structural qualities. These findings have implications for both
  software practice and research.
  Software practitioners can emphasize current quality practices as these
  influence the code that will be developed in the future.
  Researchers in the field may replicate and extend the study to improve our understanding of the theory and its practical implications on artifacts, processes, and people.
\end{abstract}


\section{Introduction}
In the
late 1960s Stanford professor Philip Zimbardo and his research team
ran a fascinating field study demonstrating the ecological effects of community
and anonymity on vandalism~\cite{Zim06}.
They removed the license plates from two used cars and abandoned them on the
street with the hood slightly raised:
one in leafy Palo Alto, California and one in New York City's gritty Bronx.
Within two days they recorded 23 instances where people tore apart or wrecked
the Bronx car.
In contrast, in Palo Alto in a five day period the only person who touched
the car was a passerby who on a rainy day caringly closed the hood to
protect the motor.
In his description of the experiment
Zimbardo argues that in environments where anonymity and a lack of community
sense are the rule,
individuals resort to vandalism and graffiti to gain personal recognition.

In 1982 the academic George Kelling and James Wilson used
a news report of Zimbardo's demonstration~\cite{Diary69},
together with an evaluation of New Jersey's police foot-patrol program
and their personal observations of Newark foot-patrol officers,
to discuss policies for maintaining safe communities.
In a long, influential, and somewhat controversial article,
titled {\em Broken Windows}~\cite{KW82}, they argued that
the maintenance of public order can
lead to safer communities.

The views of Kelling and Wilson, termed as the {\em broken windows} theory,
have been used
to explain the variation of crime among neighbourhoods~\cite{Sko92},
support theories linking disorder with crime~\cite[pp. 281--281]{HL06},
and set public policy,
most famously in the 1990s by William Bratton as Rudy Giuliani's
New York City police commissioner~\cite[pp. 47--50]{Har04}.
There is no agreement on the results of the corresponding
policies~\cite{Sko92,KC97,Har04,HL06}, mainly because it is difficult to
perform controlled studies on the subject.
A large carefully-controlled study of Chicago neighbourhoods
found that social cohesion among neighbors
and their willingness to intervene for the common good is associated with
reduced violence~\cite{SRE97}.
More recently, six clever field experiments demonstrated that when
people are exposed to the violation of observed (descriptive)
social norms and rules,
they are significantly more likely to break prescribed (injunctive)
norms and rules~\cite{KLS08}.
On the other hand,
a study published in the same decade~\cite{HL06} independently
recreated and examined Kelling and Wilson's data and
attributed the original attributed New York's
crime reduction to mean reversion.
The same study also examined a randomized social experiment
that moved families to less disorderly neighbourhoods.
The study failed to find a corresponding reduction in
those people's criminal behavior.
Furthermore,
a more recent meta-analysis of 96 studies~\cite{BFW19a}
failed to find consistent evidence that disorder increases aggression
or deteriorated attitudes toward the neighborhood.
Finally, a meta-analysis  of 198 studies by the same authors~\cite{BFW19b}
identified methodological weaknesses that have inflated evidence for
the broken windows theory and
found an association from disorder
to lower mental health,
but not to physical health or risky behavior.

Despite these mixed findings in social contexts,
the concept of the broken windows theory has intrigued researchers in various fields, including software development. 
The \textit{objective} of this study is to investigate the
broken windows theory in the context of software development,
\ie{}  examine whether
developers become more or less diligent regarding their coding,
depending on the internal quality of the code they operate on.
Internal quality comprises the aspects of
software quality that are experienced mainly by its developers
rather than its users.
It includes the code's formatting, structure, and identifier naming.
On the other hand, internal code quality does not cover the software's
functionality, reliability, or performance;
the things that are often the topics of defect or bug reports.
In an analogy to the broken windows theory, we consider
code of high internal quality as ``order''
and changes that reduce it as ``crime''.
There is also an analogy with the seriousness of the crime:
code style infractions~\cite{Spi11a}
can be considered as petty crime,
whereas structural problems are more serious.

Though known and often anecdotally referenced, 
the broken windows theory has not been explored adequately in our field.
A motivation for the study is to justify devoting
effort to maintaining specific attributes of internal code quality
because of their indirect, signalling, effects.
The findings can have implications regarding the software development process
in general and also specific aspects such as tooling, refactoring,
code reviews, and continuous integration.

This work,
based on the statistical analysis of
metrics derived from two million code commits
in 122 constantly evolving projects for long time,
comprising 5.5 million lines of code ({\sc loc})
contributes two types of findings.
First, history's weight on the evolution of internal code quality:
it seems that some aspects of a body's existing code quality are related to
the quality of its subsequent evolution.
Second, the relationship between the commits' code quality in areas covered
by injunctive norms and some of the corresponding code's descriptive norms:
adherence to coding guidelines is related to the look of the existing code.
Both findings provide (qualified) support for the application of the
broken windows theory to software development.

We contribute the following to the state of the art.
First, we present a method to systematically explore
the applicability of the broken windows theory on code quality.
Next, we outline a theoretical model
concerning code quality and the broken windows theory,
contributing towards the understanding of their potential connection.
Finally, we make publicly available a replication package
comprising time-series code quality data (metrics and smells)
from 122 open-source projects as well as corresponding analysis
scripts.\footnote{\url{https://doi.org/10.5281/zenodo.13142720}\label{fref:replication}}


\section{Theoretical Model} 
\label{sec:theoretical-model}
The impact of norms on human behavior can be productively studied by
distinguishing two norm types.
{\em Injunctive} norms encompass what others approve or disapprove,
in a formal way (through rules) or informally (through social pressure).
{\em Descriptive} norms illustrate what others actually do~\cite{CRK90},
and are established when a subject observes the environment.
Both norms provide information, what is the expected and the common behavior,
and in a particular situation they can be in agreement or in conflict.
In the context of programming, an injunctive norm would be
the disapproval of using the {\em goto} statement~\cite{Dij68b},
while a (conflicting) descriptive norm would be its common
use to jump to a function's error exit routine
\cite{GKP96}, \cite[pp. 43--44]{Spi03i}.
In the physical world, it has been found that injunctive norms are
observed more when they are in agreement with descriptive norms,
associated with the corresponding or even another type of
behavior~\cite{CG04,KLS08}.
Persons take into account these norms in order
to accurately model reality and their reactions,
to have meaningful social relationships, and
to maintain their self-concept~\cite{CG04}.

For the purposes of our study we have devised a model
that describes our understanding of the factors affecting
the evolution of a code module's internal quality from a time point $T_N$
to a later time point $T_{N+1}$.
First, there are  factors external to our study that affect the
quality at both time points, without however establishing a causal
correlation via the code.
These may include developer ability, the development process
(which can set diverse injunctive norms through rules, guidelines, processes,
and tools),
the application domain, the global distribution of developers,
and many others; see Section~\ref{sec:rel}.
All these can affect the code quality at both time points,
and thereby result in a correlation.
There are also factors that may establish a direct causal correlation
between the code at the two time points,
\ie{} the code quality at $T_N$ directly affecting the code quality at $T_{N+1}$.
The most important factor is clearly the legacy of inherited code:
at time point $T_{N+1}$ the majority of the code,
and therefore its quality,
will consist of code from $T_N$ with some changes.

There are also other more interesting ways in which the code quality at $T_N$
affects the code quality at $T_{N+1}$.
First, come the interfaces that the code has to use,
either to interface with the rest of the system
or to use third party components.
If their design is substandard, they can be detrimental to the
code quality~\cite{Spi98a,BW16,MKAM18},
because they can impose naming convention violations or,
worse, an ineffective module structure.
Then, comes the module's design structure.
If this imposes bad traits,
such as lack of appropriate layering or encapsulation,
subsequent code additions that build upon that design
will naturally add to the problem.
Also consider {identifier naming}.
Badly named identifiers
(overly short, long, inaccurate, or violating coding conventions)
existing at $T_N$ are likely to be used at $T_{N+1}$,
thus perpetuating the problem.
Finally, any of the preceding aspects of the internal code quality
may act as a descriptive norm,
signalling to its developers a lack of care regarding a module's quality,
contributing to the commission of further sins in the future.
This last point is the essence of the broken windows theory applied
to software development.

Although our examples showed how bad code can lead to worse,
note that all the factors associated with the code
can also improve the code quality:
a sound design, suitably-named identifiers, high-quality third party components,
and a code state that signals love and care can make subsequent additions
maintain or improve the code's quality.
Also note that, although the module structure, the naming conventions,
and the used interfaces at $T_{N+1}$ are bound by those used
at point $T_N$, this binding can be broken through refactoring~\cite{Fow00}.
However, in practice, refactoring tools,
which can aid these tasks, are underused~\cite{MPB12},
and it is doubtful whether refactoring actually improves
code quality metrics~\cite{SS07,Als09,SGK09,BBDO15}.

In common with the real world, the signalling effect associated
with the broken windows theory in software development
is about communicating expectations regarding community standards.
These can be associated with
{\em injunctive and descriptive norms} (this is how we write code around here),
{\em incentives}
(rewarding or reprimanding developers based on the quality of their code),
and {\em processes} that flag or correct quality problems
(code reviews, commit hooks, and continuous integration checks).

\begin{figure}[h!]
\includegraphics[width=.5\textwidth]{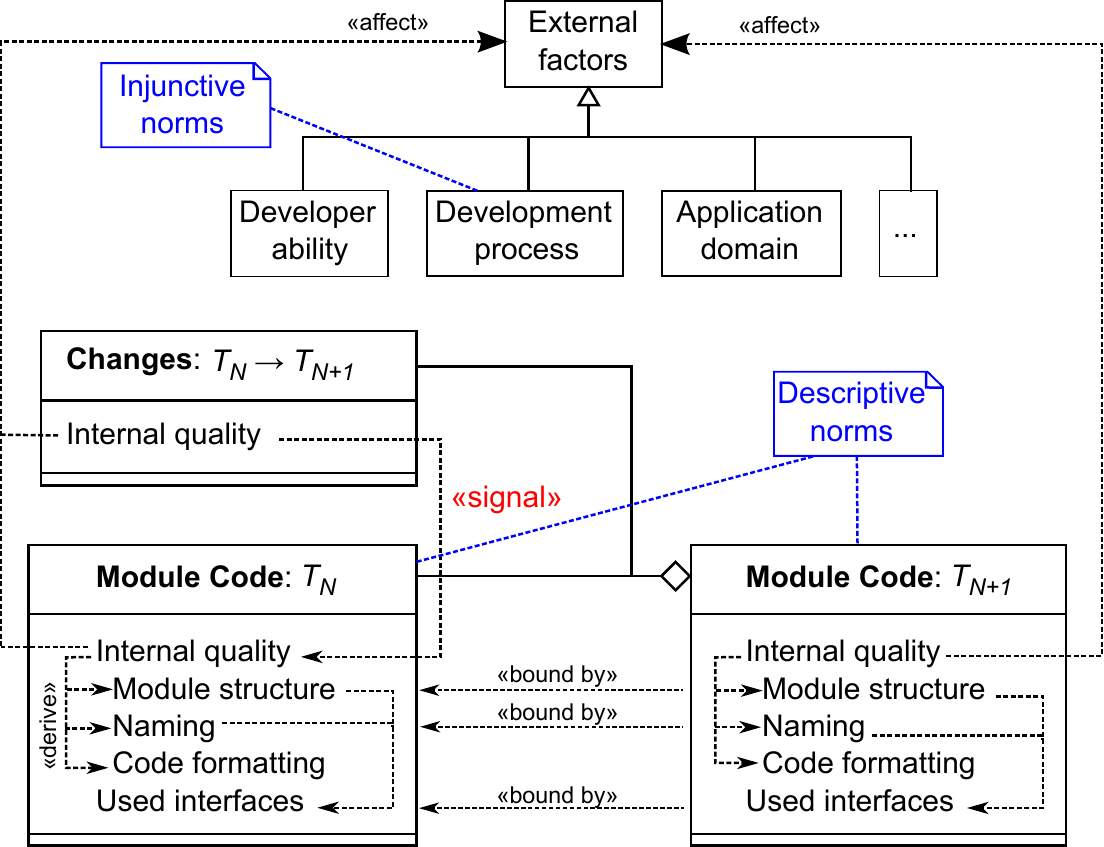}
\caption{Factors affecting the evolution of internal code quality.}
\label{fig:model}
\end{figure}

The code quality evolution model we use in our study
is illustrated as a {\sc uml} diagram in Figure~\ref{fig:model}.
Note that {\sc uml}, confusingly for some, has dependency arrows point from the
dependent (client) entity to the independent (supplier) one.
So in the figure the internal quality of the changes $T_N \rightarrow T_{N+1}$
may depend on the signalling effect of the module's internal quality at $T_N$;
the internal quality is derived from the module structure, naming,
and formatting;
the external factors affect the internal quality;
and the structure naming and interfaces at $T_{N+1}$ are bound by the choices
at $T_N$.

\section{Methods}
\subsection{Overview}
The goal of this study is to examine the applicability of the broken windows theory to source code quality.
Specifically, we investigate whether the adherence to code quality practices within a source code project impacts its internal quality.
In other words, we explore whether the historical internal quality of the code influences developers' adherence to code quality practices.
To achieve this goal, we formulate the following two research questions.

\begin{itemize}
	\item
	[\textbf{RQ1.}]
	\textit{Does future code quality relate to past code quality?}
\end{itemize}
	Through this research question we investigate whether the code quality 
	at time $T_N$ correlates to the code quality at time $T_{N+1}$.
\begin{itemize}
\item [\textbf{RQ2.}] \textit{How does existing code quality relate to
    developers’ behavior and practices concerning it?}
	\end{itemize}
	With this research question
	we check whether a developer behaves better or worse
	on source code that starts with better or worse quality characteristics.
	Exploring these research questions will help us establish the
        applicability (or, non-applicability) of the broken windows
        theory on code quality.

To investigate the research questions,
we conduct an empirical analysis.
First, we identify a set of repositories written mainly in C and Java.
We then collect a time series of
the required code quality metrics and code smells indicating
the code quality of the repositories.
We apply statistical analysis to the collected code quality data
to investigate the addressed research questions.
The rest of the section describes each of the steps
in detail.


\subsection{Data Collection and Processing} 
\label{sec:processing}

We studied the possible correlation between the quality of an existing
code body and additions or changes made to it as follows.
After selecting two popular languages exhibiting different aspects
of code quality characteristics (C and Java), we employed stratified
sampling to obtain a random representative sample of GitHub open source code
repositories to study.
We selected various metrics and smells covering size, 
structure, code style, documentation,
and adherence to design principles.
These metrics and smells are commonly used in
code quality analysis studies and fairly
represent the state of software code quality.
We chose two tools that can reliably produce quality measures for
diverse projects written in these languages,
namely \cmcalc~\cite{SLK16} for C and
\textsc{DesigniteJava}~\cite{DesigniteJava} for Java.
Based on the capabilities of these tools, we then collected data regarding
the evolution of C code quality metrics and Java design, implementation,
and testability smells for more than two million file revisions.
Finally, we analyzed the obtained data
using statistical autocorrelation techniques.
We have made publicly available on the Zenodo repository
a 260 MB replication package with
the code used for extracting and analyzing the data as well as the obtained
results.\footref{fref:replication}

\subsubsection{Project Sampling} 
\begin{table*}[tb]
\caption{Examined Repository Metrics}
\label{tab:repos}
\centering
\scalebox{1}{
\begin{tabular}{lrrrrrr}
 & \textbf{Total} & \textbf{Min} & \textbf{Median} & \textbf{Avg} & \textbf{Max} & $s$ \\
\toprule
Created (YYYY-MM)	& 	& 2008-07	& 2012-02	& 2011-11	& 2013-08	& 16 \\
Commits	& 2\,233\,372	& 223	& 3\,624	& 18\,306	& 872\,689	& 80\,321 \\
Committers	& 	& 13	& 157	& 478	& 24\,960	& 2\,263 \\
Stars	& 	& 26	& 6\,628	& 13\,787	& 174\,980	& 21\,767 \\
Forks	& 	& 22	& 1\,451	& 4\,072	& 52\,625	& 7\,889 \\
Files	& 386\,909	& 35	& 966	& 3\,171	& 65\,697	& 7\,009 \\
Lines	& 104\,442\,209	& 3\,589	& 164\,919	& 856\,084	& 27\,533\,925	& 2\,640\,579 \\
C files	& 56\,389	& 3	& 183	& 842	& 27\,626	& 3\,373 \\
C lines	& 35\,306\,678	& 6\,475	& 92\,137	& 526\,965	& 18\,904\,362	& 2\,302\,884 \\
C file revisions	& 2\,195\,510	& 38	& 3\,884	& 32\,769	& 1\,148\,732	& 140\,756 \\
C analysis time (s)	& 145\,644	& 1	& 38	& 2\,174	& 132\,335	& 16\,148 \\
Java files	& 82\,090	& 6	& 628	& 1\,493	& 12\,568	& 2\,295 \\
Java lines	& 12\,378\,220	& 137	& 81\,294	& 225\,059	& 2\,179\,378	& 392\,288 \\
Java file revisions	& 638\,533	& 4	& 6\,510	& 11\,610	& 50\,728	& 13\,534 \\
Java analysis time (s)	& 1\,599\,688	& 303	& 4\,610	& 34\,036	& 507\,184	& 89\,989 \\

\end{tabular}
}
\end{table*}

Following proposed guidelines for the systematic mining of software
repositories~\cite{Vid22,DAB21} we established the following
{\em inclusion criteria} for selecting repositories.
\begin{enumerate}
\item
More than 10 GitHub stars or forks, to select projects relevant to the
software engineering community, and avoid personal projects
and student exercises.
These two metrics are considered to be the most useful GitHub project
popularity metrics~\cite{BV18}.
\item Code in the C or the Java programming language.
These two languages
cover the imperative and object-oriented programming paradigms,
and are among the most popular languages according to the \textsc{TIOBE} index.%
\footnote{\url{https://www.tiobe.com/tiobe-index/}}
\item
At least ten years of history with at least one commit on every
half-year interval, in order to examine long-evolving
software, where code quality may act as a repository of tacit norms.

\end{enumerate}

For {\em repository selection} we employed {\em random sampling},
aiming for a sample size $N$ in the range 50--100,
which is what we could process with the computational and storage resources
at our disposal.
To aid the generalizability of our findings,
we selected a random stratified sample of the $N$ projects
based on community interest and third-party involvement.
Specifically,
we defined our sampling method to reflect the fact that
it is more likely for developers to interact with popular projects.

In common with most studies, we sampled projects from
GitHub, which contains millions of open source repositories,
including mirrors of popular projects hosted elsewhere.
We used GitHub stars as a proxy for community interest~\cite{BV18} and
GitHub forks as a proxy for direct developer involvement.
We defined five strata on an exponential progression of star or fork
engagement ranges:
11--100,
101--1\,000,
1\,001--10\,000,
10\,001--100\,000,
100\,001--1\,000\,000.
(Through experiments we determined that this stratification yields same
order of magnitude number of projects in the first four strata.
We also found that there are no projects with more than 1\,000\,000
stars or forks that satisfied our other criteria.)

We calculated the number of projects to sample in each stratum
to mirror the probability of engaging with the stratum's projects as follows.
For each stratum we obtained with GitHub \textsc{api} queries
the number of projects in it that were written in one of the languages
we studied, and had at least ten years of history,
including a change in a six-month period in the period's middle.
In each stratum $i$ for the number of obtained projects $P_i$ in it,
we estimated the total engagements (forks or stars) $T_i$ in the range
$10^i$ to $10^{i+1}$ as
\[
T_i= P_i \frac{10^{i} + 10^{i + 1}}{2}
\]
Based on it, we calculated a random selection probability to obtain
one of the $N$ projects if it had a single engagement as
\[
S=\frac{N}{\sum T_i}
\]
Finally, we obtained the number of projects $N_i$ to select in each stratum $i$
as $N_i = S T_i$.
For example,
in the Java projects stratum with
10\,001--100\,000 stars there are 51 projects,
sharing an estimated total of 2.8 million stars,
which results in the requirement to select 30 projects from it.
We then selected projects at random from each stratum $i$,
checked whether they satisfied the outlined criteria,
and kept those that did,
until we reached the required number of projects $N_i$.
Doing this for both stars and forks and taking into account
overlaps yielded a number of projects in the desired range 50--100:
93 for Java and 83 for C.

We obtained all data using GitHub \textsc{api} queries, utilizing features such as
range selection, sorting,
and the conjunction of multiple selection criteria in a single query
to obtain the required results in the most efficient manner.
This required 36 queries for obtaining the strata metrics
(6 strata $\times$ 2 engagement metrics $\times$ 3 languages
--- we also provide C\# projects in the dataset to facilitate future work)
and more than 6\,000
for deriving the projects sample
(18 queried commit intervals for 304 accepted projects in all three languages,
plus average of 9 intervals for 139 rejected projects).
As a last step we we examined the list of randomly selected repositories
for potential issues and removed two repositories
that were clones of the (also selected) Linux kernel
({\em Xilinx/linux-xlnx} and {\em raspberrypi/linux})
and one that contained mostly patches rather than code (freebsd/pkg).
Indicative descriptive statistics of the examined project repositories
are listed in Table~\ref{tab:repos}.
The file and line metrics refer to the contents of each repository at the
head of its default branch.

\subsubsection{C Quality Metrics Collection} 
\begin{table*}[!t]
\caption{C Code Size and Quality Metrics per File}
\label{tab:sizemetrics}
\centering
\begin{tabular}{llrrrrrrrr}
\multicolumn{2}{l}{\textbf{Metric and initials}} & \multicolumn{2}{c}{\textbf{Mean}} & \multicolumn{2}{c}{\textbf{25\%}} & \multicolumn{2}{c}{\textbf{50\%}} & \multicolumn{2}{c}{\textbf{75\%}} \\
& & \textbf{Begin} & \textbf{End}
& \textbf{Begin} & \textbf{End}
& \textbf{Begin} & \textbf{End}
& \textbf{Begin} & \textbf{End} \\
\toprule
Number of statements	& 	& 119.8	& 168.4	& 11	& 14	& 49	& 65	& 133	& 181 \\
Number of characters	& 	& 13079	& 18568	& 2488	& 2783	& 5748	& 7245	& 13085	& 17749 \\
Number of comment characters	& 	& 2551	& 3382	& 435	& 409	& 1101	& 1159	& 2469	& 2823 \\
Number of comments	& 	& 30.76	& 39.72	& 2	& 3	& 8	& 11	& 25	& 33 \\
Comment density \%	& CD	& 67.46	& 57.85	& 6.383	& 7.692	& 14.68	& 15.44	& 29.89	& 30.49 \\
Comment size	& CS	& 191.2	& 154.4	& 60.47	& 53.88	& 107.4	& 88.8	& 208	& 157 \\
Number of functions	& FN	& 10.48	& 14.27	& 1	& 2	& 5	& 6	& 12	& 16 \\
Function size	& FS	& 12.59	& 13.05	& 6	& 6.333	& 9.75	& 10.06	& 14.71	& 15.09 \\
Goto density \%	& GD	& 1.689	& 1.863	& 0	& 0	& 0	& 0	& 1.695	& 2.128 \\
Mean unique identifier length	& IL	& 10.2	& 10.71	& 8.374	& 8.915	& 10.04	& 10.6	& 11.75	& 12.31 \\
Mean line length	& LL	& 27.24	& 26.73	& 22.67	& 22.98	& 25.39	& 25.57	& 29.04	& 29.3 \\
Number of lines	& LN	& 445.5	& 619	& 90	& 101	& 212	& 266	& 474	& 635 \\
Questionable word density \%	& QD	& 0.1162	& 0.08946	& 0	& 0	& 0	& 0	& 0	& 0 \\
Style inconsistency \%	& SI	& 2.075	& 1.767	& 0.09494	& 0.1293	& 0.9804	& 0.8386	& 2.708	& 2.254 \\
Mean statement nesting	& SN	& 0.5738	& 0.5916	& 0.2727	& 0.3023	& 0.4945	& 0.5188	& 0.7679	& 0.781 \\

\end{tabular}
\end{table*}

Calculating quality metrics on large C code bodies is tricky
for technical and
operational reasons~\cite{Moc09,GS13}.
On the technical side, dependencies of code associated with
its compilation environment,
as well as code portability issues,
make it difficult to
establish the context required to parse and
semantically analyze the code.
This is especially true for programs written in C,
where the compilation depends
on compile-time flags and macro definitions passed through the build process,
system header files,
the search paths for these files, and
macros internally defined by the compiler~\cite{Spi03r,LKA11,GG12}.
Then comes the required throughput:
with build times for large projects taking minutes,
analyzing thousands of revisions of hundreds of projects
with a full-fledged compile can take an impractically long time.

We addressed both problems choosing to use,
in common with other studies~\cite{SLK15,Joh16,Wal20,Wal22},
\cmcalc,
an open source tool that efficiently calculates C code quality metrics,
without requiring full access to the compilation environment's
parameters~\cite{SLK16}.
The \cmcalc\ tool operates as a filter,
receiving on its standard input C source code,
and printing on its standard output a line of metrics associated with
that code.
As such it can be efficiently tied to the output of a
{\tt git show} command, so that successive versions of a file
can be analyzed without the performance degradation of
code touching the secondary storage.
The tool's operation is based on a state machine logic lexical analyzer
for a superset of C code.
The analyser combines the functionality of the C preprocessor
and C language-proper lexical analysis with rudimentary parsing,
so as to recognize
C preprocessor directives,
functions,
statement nesting,
indentation,
other spacing,
comments,
identifiers,
keywords, and
operators.
The provided metrics do not require semantic analysis of the code,
allowing \cmcalc\ to sidestep its cost and brittleness;
thus \cmcalc\ dodges the complexity of C's pointer aliasing.
By treating the C preprocessor's function-like
(those defining entities that can be called like a function)
and
object-like (e.g. those defining constants)
macros as C-proper functions and objects,
\cmcalc\ will produce reasonably accurate results without requiring access to
header files and the compilation environment.
As an example, \cmcalc\ will not stop processing with an error due
to missing include files, declarations, or definitions.

The \cmcalc\ tool calculates size, structural, quality,
and code style metrics;
see references~\cite{Kan02},~\cite[pp. 326--333]{Spi06} for more details.
A representative selection of metrics, along with the quartile points calculated
on the first version and last version of each file in our data set,
are listed in Table~\ref{tab:sizemetrics}.
In our study we provided \cmcalc\ with successive versions of each
C file, and obtained from its output the corresponding metrics:
one set of metrics for each version of each C file.
Most metrics are self explanatory; here are details for the rest.
The comment density ({\sc cd}) is the ratio between the number of comments
and statements in a file.
The comment size ({\sc cs}) is the ratio between the number of comment characters and
the number of comments.
The function size ({\sc fs}) is the ratio between the number of statements and
the number of functions.
The goto density ({\sc gd}) is the ratio between the number of {\em goto} statements and
the number of statements.
The questionable word density ({\sc qd}) is the number of words
that may indicate problems in the code as well as swearwords
divided by the file's number of lines.
The following whole words were searched for in a case-insensitive manner:
bugbug,
buggy,
bullshit,
crap,
crash,
damn,
damned,
doom,
doomed,
fixme,
fuck,
fucker,
fucking,
hack,
hacked,
hackery,
hacks,
hell,
kludge,
kludges,
lame,
lameness,
poop,
screwed,
screws,
shit,
shits,
suck,
sucks,
todo,
xxx.

Finally,
the mean statement nesting ({\sc sn}) is measured as the sum of the nesting of
all lines within code blocks
(e.g.,\ 1 after a {\em while} statement and
2 after an {\em if} statement nested within the {\em while} one)
divided by the number of those lines.

The \cmcalc\ tool calculates code style infractions from commonly
agreed formatting guidelines.
As there are a number of different approaches for formatting C code,
\cmcalc\ allows us to measure the {\em consistency} of their application,
rather than adherence to a specific formatting style.
Specifically, for each way to format a particular construct
(for example putting a space after the {\tt while} keyword)
\cmcalc\ counts the times $a$ the rule is applied in the one way
(e.g.,\ putting a space) and the times $b$ the rule is applied in the other way
(omitting the space).

Each metric represents the number of occurrences of the corresponding
phenomenon in a file.
\begin{itemize}
\item A space, $a$ (or a lack of it, $b$)
before or after the following tokens:
binary operator,
closing brace,
comma,
keyword,
opening brace,
opening square bracket,
semicolon,
{\tt struct} access operator.
($2 \times 2 \times 8 = 32$ metrics.)

\item A space, $a$ (or a lack of it, $b$):
before a closing bracket,
after a unary operator,
before a closing square bracket.
($2 \times 3 = 6$ metrics.)
Note that the rules regarding spacing on the opposite side of the
preceding three tokens are context-specific,
and therefore they were not checked.

\item A space, $a$ at end of a line.
(One metric; no style convention puts a space at the end of a line,
therefore $b=0$ in this case.)
\end{itemize}

The file's style inconsistency for $n$ style rules (20 in our case)
as a percentage of possible cases is defined as follows.
\begin{equation}
{\rm SI} = \frac{\sum\limits_{i=1}^{n}\min(a_i, b_i)}{\sum\limits_{i=1}^{n}a_i + b_i} \times 100
\end{equation}
Thus, through \cmcalc\ and the preceding definition we
identify the prevalent coding style used in each file
(e.g. putting spaces around a binary operator),
and obtain a metric of inconsistency regarding the coding style
found within the file.
We obtained the rules from the
Google,\footnote{\url{https://google.github.io/styleguide/cppguide.html}}
Free{\sc bsd},\footnote{\url{http://www.freebsd.org/cgi/man.cgi?query=style&sektion=9}}
{\sc gnu},\footnote{\url{https://www.gnu.org/prep/standards/html_node/Formatting.html}} and
the updated Indian Hill\footnote{\url{https://www2.cs.arizona.edu/~mccann/cstyle.html}}
style guidelines.

\ifx\fullprint\relax
We did not use all metrics calculated by the \cmcalc\ tool.
Following a method established in an earlier study using the same
tool~\cite{Spi08b},
we planned to utilize the use of risky C preprocessor statements
as a measure of quality.
However, we did not find sufficiently large sets of developer variation
in their use (see Section~\ref{sec:results}) to allow the measure's adoption.
Furthermore,
we decided not to use the generated Halstead and cyclomatic complexity
as quality measures,
because they have not been found particularly useful as an indicator
for the code's fault-proneness~\cite{FO00}.
\fi

From the metrics we gathered five are associated with code style quality:
commenting ({\sc cd, cs}), naming ({\sc il}), and layout ({\sc si, ll}).
Another six are proxies for code structure quality:
modularity at the file level ({\sc fn, sn, ln}) and the
function level ({\sc fs}), code complexity ({\sc sn, gd}),
and questionable coding practices ({\sc qd}).

The metrics calculation was performed through a series of nested loops,
expressed as {\em bash}~\cite{New09} shell scripts
that run \cmcalc\ for each revision, of each file, of each repository.
The revisions were obtained through the {\tt git log} command,
run with a custom output format to obtain the revision's hash code,
committer email, and machine-readable time stamp.
Then {\tt git show} was invoked on the filename and hash code associated
with each revision to pipe to \cmcalc\ the source code to be analyzed.
Thus a single 118-field line was produced for each file's revision
%
(258 million values in total),
which other programs could use to analyze the results.

The metrics calculation of all revisions of all files (35 million lines)
took more than 40 hours to run.
Caching and checkpointing were used to allow the efficient execution of
incremental runs while the processing code was debugged~\cite{Spi24f}.
A considerable speedup was achieved by parallelising the analysis
of each file using {\sc gnu} parallel~\cite{Tan11}.
This gave the calculation a throughput of 25 thousand lines per second
on an 8-core computer.

\subsubsection{Java Code Smells Collection} 
\begin{table}
\caption{Java Code Size and Smell Instances}
\label{tab:javasmells}
\centering
\begin{tabular}{llrr}
\textbf{Type} & \textbf{Name} & \textbf{Begin} & \textbf{End} \\
\toprule
	& Lines of code	& 5\,872\,257	& 6\,321\,769 \\
	& Number of classes	& 143\,850	& 147\,279 \\
DS	& Broken hierarchy	& 10\,935	& 11\,183 \\
DS	& Broken modularization	& 3\,721	& 4\,111 \\
DS	& Deep hierarchy	& 13	& 14 \\
DS	& Deficient encapsulation	& 12\,719	& 13\,706 \\
DS	& Feature envy	& 5\,504	& 5\,994 \\
DS	& Hublike modularization	& 97	& 153 \\
DS	& Insufficient modularization	& 6\,840	& 8\,092 \\
DS	& Missing hierarchy	& 245	& 311 \\
DS	& Multifaceted abstraction	& 181	& 214 \\
DS	& Multipath hierarchy	& 185	& 201 \\
DS	& Wide hierarchy	& 314	& 371 \\
IS	& Complex conditional	& 6\,788	& 8\,450 \\
IS	& Complex method	& 11\,869	& 14\,188 \\
IS	& Empty catch clause	& 14\,841	& 16\,345 \\
IS	& Long method	& 1\,261	& 1\,575 \\
IS	& Long parameter list	& 14\,506	& 17\,419 \\
IS	& Long statement	& 118\,374	& 138\,959 \\
IS	& Magic number	& 301\,830	& 334\,784 \\
IS	& Missing default	& 3\,644	& 4\,110 \\
TS	& Excessive dependency	& 7\,479	& 8\,773 \\
TS	& Global state	& 12\,310	& 13\,753 \\
TS	& Hard-wired dependency	& 204	& 234 \\
TS	& Law of Demeter violation	& 437	& 628 \\

\end{tabular}
\end{table}

We extended the code quality analysis by detecting and analyzing 
commonly occurring code smells~\cite{Fowler1999, Sharma2018}
on projects mainly written in Java.
For a comprehensive coverage of code quality,
we required a tool that supports code smell detection
at different granularities
such as implementation, design, and testability.
We chose \textsc{DesigniteJava}~\cite{DesigniteJava}, which
detects a variety of code smells and computes common code quality metrics.
The tool has been validated by its authors~\cite{Sharma2020, Sharma2023}
and has been used
in diverse studies~\cite{Oizumi2019, Sharma2020, Eposhi2019, Uchoa2020, Alenezi2018}.

We included in our analysis three types of smells listed
in Table~\ref{tab:javasmells}:
design smells (DS) ~\cite{Suryanarayana2014},
implementation smells (IS)~\cite{Fowler1999}, and
testability smells (TS)~\cite{Sharma2023}.
The latter,
taking into account that
testability is the degree to which the development of test cases
can be facilitated by the software design choices,
are the practices that impact the testability of a software system.
We selected the listed commonly occurring smells for our analysis because,
given their scope and characteristics
they may get influenced by other existing smells.

In terms of its architectural design, \textsc{DesigniteJava} is organized into three layers.
At the bottom layer lies
the Eclipse Java Development Toolkit (\textsc{jdt}).
The tool utilizes \textsc{jdt} to parse the source code,
prepare \textsc{ast}s,
and resolve symbols \textit{\ie{}} associate type information with variable declarations.
The middle layer, \textit{source model},
maintains a source code model
created from the extracted information from an \textsc{ast} with the help of \textsc{jdt}.
The business logic \textit{\ie{}}
the smell detection and code quality metrics computation logic resides in the top layer.
The layer accesses the source model, identifies smells, computes metrics, and outputs the inferred information in either .\textsc{csv} or .\textsc{xml} files.
As its user, we utilized the tool
for each repository selected for analysis
in \textit{multi-commit} analysis mode.\footnote{DesigniteJava documentation---\url{https://www.designite-tools.com/docs/commands.html}}
In this mode
the tool takes the path of a Git repository as input,
switches the repository to a commit,
analyzes the code, and produces a set of \textsc{.csv} files containing smells and metrics data for the commit.
This produces  \textsc{csv} files with smell details for each file at each commit.
We subsequently filtered these to create timeline series specific to changed
files.
The total number of smells detected at the beginning and at the end of all
time series is listed in Table~\ref{tab:javasmells}.
(The total number of lines in Table~\ref{tab:javasmells} is larger than the
one shown for Java files in Table~\ref{tab:repos},
because a file can appear in many time series as it moves around
the repository.)
Our processing excludes a few Java projects with faulty repositories
({\em apache/camel})
or with processing requirements that exceeded our computing resources
(e.g. {checkstyle/checkstyle}, which run out of heap space despite
getting allocated 40 GB of \textsc{ram}).

\section{Results} 
\label{sec:results}
We present our observations corresponding to both research questions
exploring the broken windows theory in software using our corpus.


\subsection{\textbf{RQ1}: Relationship of quality of existing code on
  code quality evolution}
To set the scene we inquired into the way code quality evolves over
time; in particular, the way existing code quality relates to future
quality. After all, if existing quality does not relate to future
quality in a significant way, it does not make much sense to check the
effect of existing code in developers' coding practices.

\begin{figure}[tb]
\includegraphics[width=.4\textwidth]{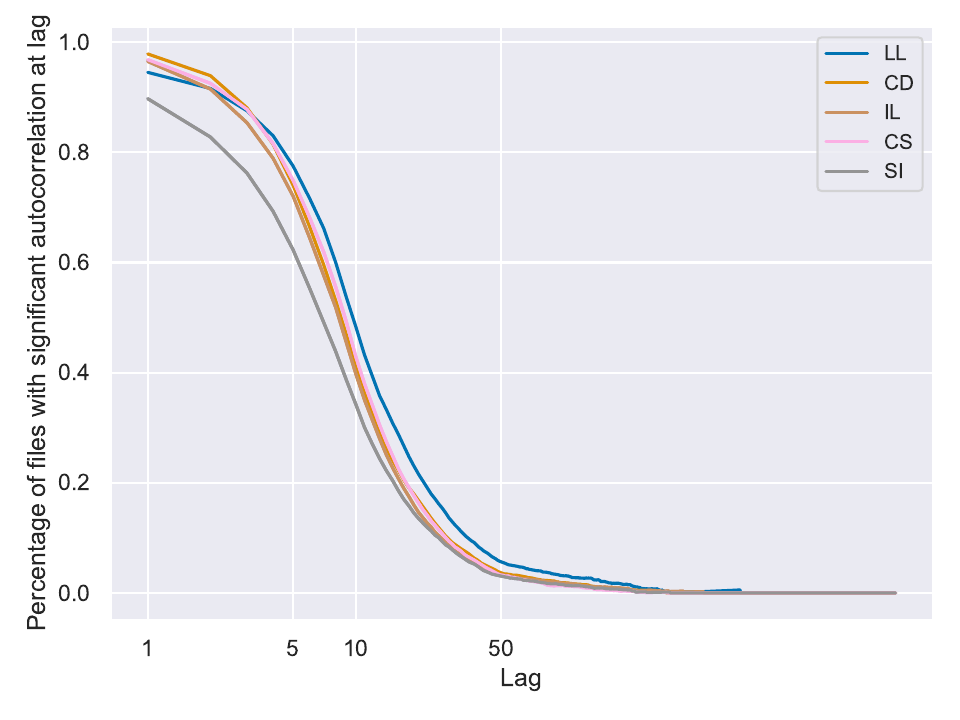}
\caption{Percentage of files with autocorrelation $>0.5$ at each lag
  for code style metrics.}
\label{fig:autocorrelation-style}
\end{figure}

\begin{figure}[tb]
\includegraphics[width=.4\textwidth]{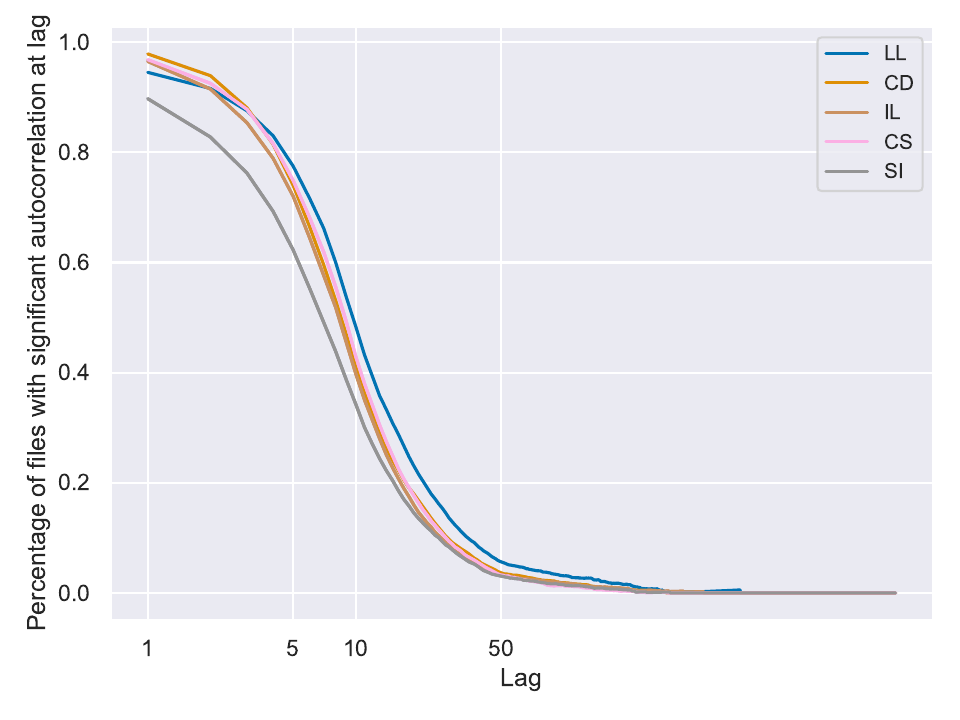}
\caption{Percentage of files with autocorrelation $>0.5$ at each lag
  for code structure metrics.}
\label{fig:autocorrelation-structure}
\end{figure}

We started our analysis by looking at the autocorrelation of C code
style metrics and C code structure metrics. We calculated the
autocorrelation for files that have more than 50 commits and a
non-constant value of the metric (because otherwise the
autocorrelation is undefined). From 55\,523 files, that left us with
10\,530 files with autocorrelations calculated for at least one of the
code style metrics and 10\,404 files for code structure metrics. We
took into account up to 50 lags and judged as important
autocorrelations greater than 0.5 that were found to be statistically
significant, having $p$-value $< 0.05$ for the Ljung-Box test. The
results can be seen in Figure~\ref{fig:autocorrelation-style} and
Figure~\ref{fig:autocorrelation-structure}.

For both classes of metrics, we see that a very high percentage of
files have significant autocorrelations for small lags. Moving forward
in time, we see that for all metrics about $40\%$ of the files have
significant autocorrelations for lags up to~10. \textbf{Overall,
  history does relate to the evolution of code style and structure
  metrics, as the value of a metric at a particular commit exerts a
  considerable effect to more than 80\% of files for all metrics in
  the commits that follow.} Moreover, in a significant portion of the
files the relationship goes deep, to many commits and not just the one
that comes next.

A general pattern that we can discern from
Figures~\ref{fig:autocorrelation-style}
and~\ref{fig:autocorrelation-structure} is that metrics that
correspond to advice given frequently in software engineering matter a
lot in history. Comments, both in what regards their density and their
size, make a mark in the history of many files. Rules on identifier
length and line length are also found in many style guides, and they
are sticky in that previous values to a large extent relate to future
values. The same goes for style inconsistency, which aggregates
different kinds of inconsistencies. Advice on function size and number
of functions in a file is frequently drilled to software developers;
so is advice against too much nesting and the use of \emph{gotos}. It
seems that it is not just history, but what we might call tradition,
in the form of old, time-honoured programming tenets, that manifests
itself along the evolution of software.

\begin{figure}[tb]
\includegraphics[width=.4\textwidth]{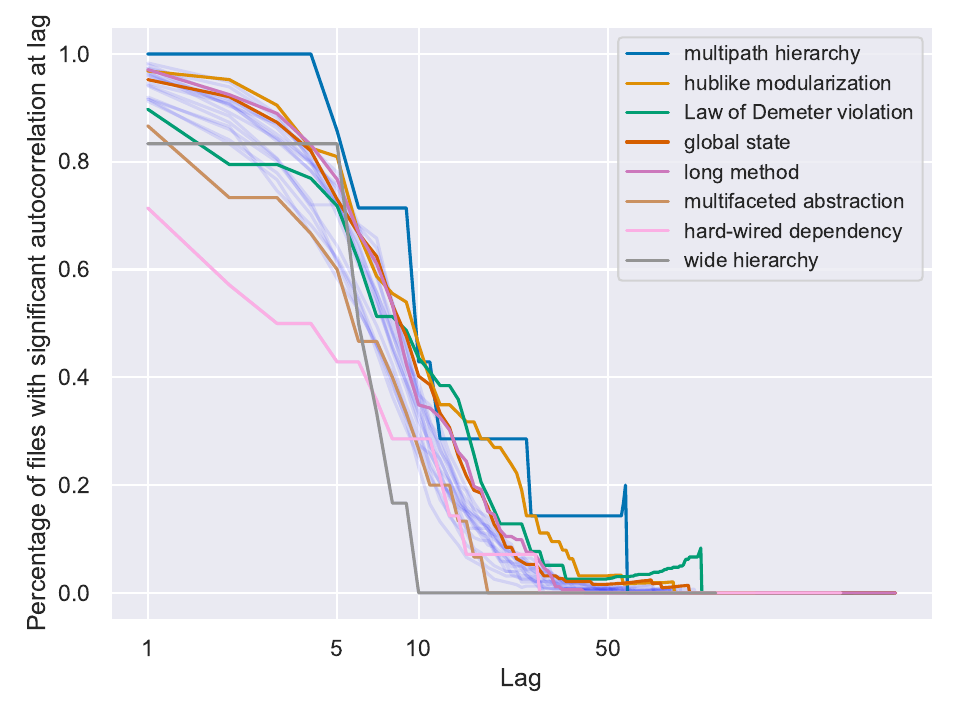}
\caption{Percentage of files with autocorrelation $>0.5$ at each lag
  for smells.}
\label{fig:autocorrelation-smells}
\end{figure}

We can discern a similar pattern in
Figure~\ref{fig:autocorrelation-smells}, which shows the
autocorrelation of Java smell metrics over different time lags. Out of
114\,519 files, we found 832 with more than 50 commits and a
non-constant metric value. The autocorrelations show again that
\textbf{about 40\% of the files we examined exhibit significant
  autocorrelations for lags up to ten}. We have highlighted the top
five and bottom three of the smells, in terms of the average
percentage. Three of the top five, \ie{} multipath hierarchy, hub-like
modularization, and Law of Demeter violation, seem to follow
distinctive paths, as do the bottom three metrics, \ie{} multi-faceted
abstraction, hard-wired dependency, and wide hierarchy. This may be
because for all the highlighted metrics, the number of files where the
autocorrelation could be calculated with statistical significance was
small: the median number of files, for the different lags, for which
we could calculate autocorrelation $> 0.5$ with $p$-value $< 0.05$ was
less than 10. 

\begin{boxH}
  Our analysis indicates that history does relate to the evolution of
  code quality, both for code style and structure metrics and for code
  smells.
\end{boxH}

\subsection{\textbf{RQ2:} Developers' behavior and historical code quality}

It is easy to spot a real broken window, but as there is no {\em a
  priori} indication of what is a good or a bad file, we used
quantiles for ascribing categories to files. For each project, we
identified the first commit and the corresponding metric for each of
the code style and the code structure metrics. We computed the 25\%
and the 75\% quantiles for those metrics and we grouped the files at
the top quantile as top files and the files at the bottom quantile as
bottom files. Note that top and bottom are not equivalent to good and
bad, as in some metrics higher values are better whereas in other
metrics the opposite is true. In total we formed 11~groups containing
top and bottom files, one for each selected metric.

To see whether developers behave differently in top files than they do
in bottom files, we investigated whether their commits in top files
are quantitatively different, in terms of our metrics, than their commits
in bottom files. In effect, we looked whether developers tailor the
quality of their commits based on the quality of the file they commit
to. We measured the quality of a particular commit as the difference
of the selected metric for the commit from the value of the metric in
the previous commit.

To work out the numbers, we grouped each project's data by developer
and for each group we selected the commits made in top files and the
commits made in bottom files. To see if the developers perform
different kinds of commits depending on the file's quality (top or
bottom), we calculated the Kolmogorov-Smirnov two samples test for
those developers that had at least 10 commits in top files and 10
commits in bottom files. We have 11 metrics, so in total we have
121~cases. As these are multiple tests, we used a Family-Wise Error
Rate of $0.05$ to adjust the $p$-values using the Benjamini-Hochberg
prodecure~\cite{benjamini:1995}.

\begin{figure}
\includegraphics[width=.4\textwidth]{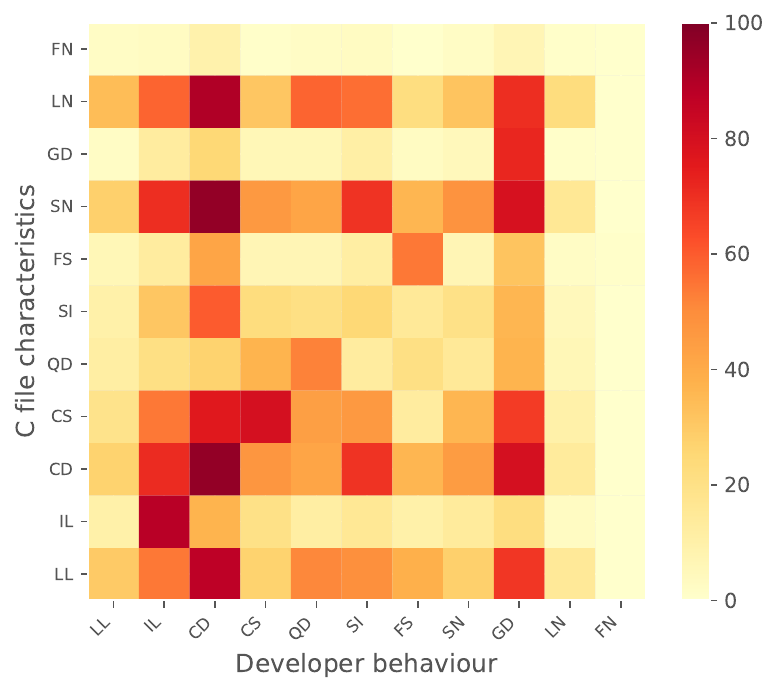}
\caption{Percentage of developers with different behaviour in top vs
  bottom C files.}
\label{fig:ks-heatmap-c}
\vspace{-2mm}
\end{figure}

The results for C code are shown as a heat map in
Figure~\ref{fig:ks-heatmap-c}. The $y$-axis of the heat map represents
the different groupings; the $x$-axis represents the metric we are
using for the statistical test. For example, the bottom left square
corresponds to files being grouped to top and bottom quartiles
depending on mean line length ($y$-axis), and the percentage of
developers that display different commit behaviour in what regards
mean line length ($x$-axis). The square at the intersection of {\sc
  si} on the $x$-axis and {\sc cd} on the $y$-axis corresponds to the
percentage of developers that display different commit behaviour in
what regards style inconsistency measured in files grouped according
to the classification of the file as top or bottom based on the code
density. In other words, what is the percentage of developers whose
commits exhibit a different style inconsistency on files coming out
top in comment density against files coming out at the bottom in
comment density?

Looking the at columns of Figure~\ref{fig:ks-heatmap-c}, {\sc cd},
{\sc gd}, and {\sc il} stand out. \textbf{Developers appear to behave
  very differently in what regards comment density, goto density, and
  to long or short identifier lengths, along files grouped in the top
  or bottom with respect to a variety of different metrics.} The
laggards are {\sc ll}, {\sc ln}, and {\sc fn}; it seems that
developers do not behave with respect to line lengths and high or low
line and function counts on files with different top or bottom metric
counts.

Going row-wise, one way to interpret each row is as the context the
file provides regarding the value of a metric in the file. \textbf{
  So, {\sc ln}, {\sc sn}, {\sc cs}, {\sc cd}, and {\sc ll} seem to
  provide particularly strong contexts and, taking the whole picture,
  stronger than others.} That is reasonable. Developers do different
kinds of commits depending on the file they are into, but not every
single characteristic of the file influences every single commit
metric to the same extent. The row for {\sc gd} illustrates that the
heatmap is not symmetric: even though developers behave differently in
terms of goto density on files grouped to the top or bottom on many
different metrics, files that come top or bottom in goto density seem
to be have high percentage of different goto density behavior on the
part of the developers.

We repeated the same analysis for the Java smells; we omit the resulting
heatmap, as it is bright yellow nearly all over. That means that we
were not able to detect a statistical significance in the Java smells
metrics of developers' commits in files rated top and bottom with
regards to those smells. That does not entail that, in contrast to
C code and structure, developers do not show a difference in their
behavior regarding design smells, implementation smells, and
testability smells; it may be that developers adhere to norms
regarding these smells independently of the context. We plan to
investigate this further in the future.

\begin{boxH}
  Our results indicate that some historical code characteristics (such
  as comment density and identifier length) strongly relate to the
  developers' behavior. However, code smells do not show a similar
  statistically significant relationship.
\end{boxH}

\section{Threats to Validity} 
\label{sec:threats}

In terms of \textbf{external validity}, it is clear that we have limited
ourselves to projects using the C and Java programming languages.
Although these languages still enjoy rude health,
programmers have a wealth of other languages at their disposal,
and it does not automatically follow that our findings transfer to them.
In our
defense, the syntax and overall style of C and Java have influenced many other
newer programming languages, so there is no {\em a priori} reason to
indicate our results would not carry over to other languages,
at least those with a similar structure.
However, it may be difficult to replicate our findings
in environments, such as Visual Basic, that by default
perform many formatting tasks automatically.
At the same time, programs written in dynamic languages may rely more
on a consistent code style to make up for the lack of a compiler
that can catch trivial errors.

The metrics we have used are internal quality ones we could
efficiently measure through the two tools we employed.
More sophisticated measures, such as the actual defect density,
may yield different results.
Looking at external quality metrics, like those measuring
reliability, accuracy, and performance, may also differentiate the
resulting picture.


In projects code style may not be a developer's choice,
but may be imposed through tools and processes.
We manually searched the developer documentation of 25 popular projects
(Blender,
CPython,
ffmpeg,
Free{\sc bsd},
\textsc{gcc},
gdb,
gecko-dev,
Git,
illumos,
ImageMagick,
{\sc kde},
Linux,
My{\sc sql} server,
Nagios,
Open{\sc cv},
Perl,
{\sc php},
Postgre{\sc sql},
{\sc vlc},
{\sc wine},
as well as the
Apache,
freedesktop.org,
{\sc gnome},
{\sc gnu} Git, and
Sourceware projects)
looking for style guidelines and how these are enforced.
We found that 7 out of 25 projects (28\%) are using automated methods
(mandatory and voluntary) to ensure code style conformance.
Linux, \textsc{gcc}, CPython, and illumos-gate
instruct committers to check their source code for style inconsistencies
before each commit,
whereas other projects suggest the use of automatic style checks,
such as adapting the configuration of source code editors and {\sc ide}s,
and executing third-party scripts.
We also found that 17 out of 25 projects (68\%)
prescribe specific mandatory or voluntary coding guidelines.
Projects with mandatory checks,
such as Free{\sc bsd}, Blender, Perl, and {\sc php},
have extensive guidelines,
and encourage committers to conform to them
in order to have their commits accepted by reviewers.
Projects with voluntary checks,
such as Postgre{\sc sql}, {\sc vlc}, and {\sc kde},
provide shorter guidelines,
and do not impose strict code style checks.

Our choice of projects may also be a limiting factor. Our random
stratified sampling resulted in the inclusion of some
big, successful projects, where quality standards may be above and
beyond those entertained by other projects. It is possible that our
findings may apply less to such other projects.
On the other hand,
it is likelier that quality in such projects will show increased variation,
and that developers will enjoy greater freedom in their coding.
Both factors would amplify the quality's signalling effects.

Note also that we only examine the changes that show up in the main
development branch. Development also takes place in other branches as
well~\cite{BRBH09};
changes on other branches may be merged with the current branch,
so these will show in the data we examine, but their own history may
be lost when commits are squashed,
as we do not examine the history of the separate branches.
This may hide from our analysis quality problems that were identified
and fixed through a code review.

Moving on to \textbf{internal validity}, we have been careful
not to propose any causal relationships in the analysis of our results,
which is a key criticism leveled against the broken windows
theory~\cite{BFW19b}.
We report results of statistical relations,
or point out the absence of relations, but
we do not attribute cause and effect. That would require a more
detailed examination of the model we proposed in
Section~\ref{sec:theoretical-model}, and, possibly, putting forward a
concrete mathematical formulation encompassing dependent and
independent variables of software quality in relation to the existing
quality context. This can be the subject of further research.

\smallskip
\section{Related Work} 
\label{sec:rel}

\noindent
\textbf{The broken windows theory in software engineering:}
To the best of our knowledge there is limited work that uses the
broken windows theory to explain factors that affect software quality.
In particular,
Deissenboeck and Pizka have referred to the broken windows theory
when examining the inconsistent naming of identifiers in software
projects~\cite{DP06}.
The same authors in another study~\cite{DWP07}
have also called for studying psychological effects,
such as those associated with the broken windows theory.
Brunet {\em et al.} refer to the theory when
suggesting that the gap between code and architecture is tractable
provided that violations are checked and solved in a short time period~\cite{BBS12}.
The preceding studies referred to the broken windows theory in order
to explain or justify software development phenomena.
However, none of the studies attempted to use empirical evidence
in order to validate or disprove the broken windows theory
in the context of software development.

A related theory concerns contagious technical debt,
which has so far been studied mainly qualitatively~\cite{MB17,FVHO23}.
In the similar vein,
the human inclination to imitate behavior of others
is known as the Bandwagon effect.
Such cognitive biases have been explored in the software engineering
context~\cite{KB2021, PR2013, PR2011, LBB22}.

\smallskip
\noindent
\textbf{Factors affecting software quality:}
On the other hand, there is a considerable body of work on the factors that
affect software quality.
These studies can be categorized into the areas of management,
the software development process,
the developers' characteristics, and
product properties.
A survey on factors that affect software quality~\cite{GL10}
examined organizational, technical, and individual factors.
In the following paragraphs we will briefly describe some representative
findings from each area.

In the field of \textit{management} a number of studies
found
that clear source code ownership results in
fewer failures~\cite{BNM11} and defects~\cite{NMB08},
that unfocused teams working on central modules can increase post-release
failures~\cite{PNM08},
that well-coordinated teams can reduce software failures~\cite{CH13},
that organizationally (rather than geographically) distributed
teams lead to an increase of software failures~\cite{BND09},
\if\isdoubleblindsub0
that geographical distance {\em per se} does not affect quality~\cite{Spi06d},
\fi
and
that low staff morale and excessive turnover can decrease software
quality~\cite{EGK01}.

Regarding the \textit{software development process}, major factors that have
been reported to affect software quality include
the software's architecture and design~\cite{EGK01,BDS98},
the quality of the requirements~\cite{KKK00,EGK01},
tool use (in some cases)~\cite{KKK00,EGK01},
test-driven development~\cite{NMBW08,JS08},
code reviews~\cite{KP09b,MKA14},
scheduling~\cite{EGK01},
(maybe) refactoring~\cite{Als09,SS07},
and the existence of software processes in general~\cite{KKK00,EGK01}.

On the \textit{developer} front, studies have found that
the capabilities and experience of the personnel can
ensure the software's conformance quality~\cite{KKK00},
while their absence can lead to software decay~\cite{EGK01}.

Finally, the properties of the developed product can also affect its
quality.
Factors that have been identified include
the product's size~\cite{KKK00,AC07,EGK01},
the age of the source code~\cite{EGK01,CH13},
code porting and reuse~\cite{EGK01,KG06},
the chosen programming language~\cite{SK03},
and the application domain~\cite{ER97,Spi08b}.

\section{Discussion and Implications} 
\label{sec:discussion}

In the experimented boundary, factors, and context,
we have shown the following in the preceding sections.

\begin{itemize}
\item The quality of an existing, initial, code body and its
  subsequent evolution are related. Code quality carries its history
  with it, and its current state is strongly dependent on that
  history. The exact nature of the dependence varies on how we define
  history. It appears that traditional software engineering advice has
  a particularly strong effect on history.
\item The developers' behavior associated with a variety of
code style and structure metrics is significantly related to a file's
commenting (comment size and density---{\sc cs, cd}),
size (number of lines and statements---{\sc ln, sn}), and,
rather unexpectedly, length of lines ({\sc ll}), 
as seen by studying Figure~\ref{fig:ks-heatmap-c} row-wise.
\item The developers' behavior associated with
identifier length ({\sc il}), comment density ({\sc cd}), and
style inconsistency ({\sc si}) is related with
a number of file characteristics, as seen by studying
Figure~\ref{fig:ks-heatmap-c} column-wise. 
\end{itemize}

In effect, we have seen evidence that some descriptive norms
that apply to software (how the software is actually written)
are associated with variations in developer behavior in areas that are often
covered by injunctive norms (guidelines and best practices).
In the context of the broken windows theory,
we did not find as high a significant relationship between
public order (style consistency)
and more serious crimes (increased statement indentation,
fewer comments, shorter identifiers, or more {\em goto} statements).
Consequently, our results are nearer to Zimbardo's original
demonstration~\cite{Diary69} ---
a community's effect on descriptive norms --
than to Kelling and Wilson's interpretation~\cite{KW82} ---
the escalation of violations from descriptive to injunctive norms.

Surprisingly, we saw that a file's style consistency
which is a ubiquitous indicator of order in a file
and thereby forms a strong descriptive norm,
is associated with the developers' behavior at a lesser extent;
certainly
not as strongly as we would have expected when we started this study.

The apparent lack of a strong behavioral link
between style inconsistency and other measures
of software quality surprised us at first;
but then upon deeper reflection less so.
The fact that it does not appear to be correlated
with other measures suggests
that it is an independent quality variable, and not one that can be
readily 
calculated from structural quality metrics.
Programming style is a distinct quality attribute, and our style
inconsistency density metric may be one way to measure it.

In the context of the broken windows theory in software development,
style inconsistency is special for a number of reasons.
First, style infractions can be easily determined by inspection,
and committed with the sure knowledge that they will not affect
the software's external quality.
Therefore, stylistic infractions are
both a more noticeable signal and a more sensitive effect.
This situation resembles the actual broken windows in the real world.

Furthermore, code style is also a matter of personal taste and opinion.
Some developers hold these opinions with such a conviction
that it may give rise to so-called holy wars~\cite[p. 35]{Goo07}.
Therefore, it may be easy for developers lacking self-discipline to
commit stylistic infractions,
especially when editing a file where their ``religion'' appears to be
tolerated.

Assuming that the effect of code on code is indeed casual, it has
important implications for software developers and their managers.
The key message that should be carried away is that
keeping basic code hygiene can not only help the software's maintainability,
but also improve the quality of subsequent code additions.
Developer behavior regarding the code quality measures we examined
can be improved by a few, rather unglamorous, actions:
keeping modules (files in the case of C code) short and focused,
writing plenty of descriptive comments, and
avoiding long code lines.

Furthermore, given that
a file's  number of functions ({\sc fn}) and their size ({\sc fs})
seem to be strongly related with
developers' behavior regarding the file's size,
it seems important to invest effort in designing the appropriate
decomposition of the code into modules
(files and functions in the case of C code), for once this structure is
set in code it is apparently difficult to escape from it.

\section{Further Work} 
\label{sec:further}
The study of social influence in the context of software development
can be expanded in two fronts:
those of the empirical data and the mechanisms at work.


The empirical basis can be extended by studying more and smaller
projects where external quality-setting factors are less prevalent.
This can be done by using data sets that contain metadata from
many such repositories, such as
the {\em RepoReapers Data Set}~\cite{MKCN16},
or {\em GitHub Search}~\cite{DAB21},
in conjunction with actual commit data from the corresponding repositories.
The work can also be easily extended to cover more programming
languages, especially given the fact that the analysis
performed by \cmcalc\ is mostly programming language agnostic.
In particular,
it would be interesting to study
object-oriented metrics~\cite{CK94},
resource leaks,
code duplication,
and security vulnerabilities.
It would also be particularly interesting to check the effect of style
infractions in languages where these often lead to errors in
programming logic;
for example, one could examine the role of the optional semicolons
in JavaScript.

Studying the mechanisms through which social influence theory applies
to software development as well as the effects of these mechanisms on
software quality and process is more challenging. Some topics worthy
of further, qualitative, examination are the following.
\begin{itemize}
\item Are the signals communicated and acted upon subconsciously,
or are developers making conscious rational choices on the quality of their
work based on a file's perceived quality?
\item Are developers using the signals to optimize where they will
direct their efforts?
\item How does this optimization affect external code quality?
\item How should the software development process be adjusted to take
  into account these factors?
\item What are the reasons and the meaning in the differences of the
  results between code and structure metrics on the one hand, and
  design, implementation, and testability smells on the other?
\end{itemize}
These are clearly questions whose answers can change the way we
view software development.

\subsection*{Acknowledgements}
This research has been co-financed by the European Union
(European Social Fund --- {\sc esf}) and Greek national funds
through the Operational Program ``Education and Lifelong Learning''
of the National Strategic Reference Framework ({\sc nsrf}) ---
Research Funding Program: Thalis ---
Athens University of Economics and Business ---
Software Engineering Research Platform;
the European Union's Horizon 2020 research and innovation programme
under grant agreement No. 825328 (FASTEN project);
and from the European Union's Horizon 2021 research and innovation programme
under grant agreement No. 101070599 (SecOPERA project).
In addition, {\sc nserc} Discovery grant (\#R35616) has also partially supported the study.

Dedicated to the loving memory of
the criminologist Professor Calliope D. Spinellis~(1933--2024).

\balance

\ifx\mypubs\relax

\else

\fi 

\end{document}